\documentstyle[aps,pre,12pt,tighten]{revtex}  

\begin{document}
\draft

\title{Random Matrix Theory and Classical Statistical Mechanics: \\
 Spin Models}

\author{H.\ Meyer\cite{byline} and J.-C.\ Angl\`es d'Auriac\cite{byline}}
\address{
Centre de Recherches sur les Tr\`es Basses Temp\'eratures,\\
BP 166, 38042 Grenoble, France}

\date{PREPRINT December 18, 1996}

\maketitle

\begin{abstract}
We present a statistical analysis of spectra of transfer matrices
of classical lattice spin models; this continues the work on the
eight-vertex model of the preceding paper.
We show that the statistical properties of these spectra can serve
as a criterion of integrability. It provides also an operational numerical
method to locate integrable varieties. In particular, we distinguish
the notions of integrability and  criticality
considering the two examples of the three-dimensional Ising
critical point and the two-dimensional three-state Potts critical point.
For complex spectra which appear frequently in the context
of transfer matrices, we show that the notion of independence of eigenvalues
for integrable models still holds.
\end{abstract}


\section{Introduction}

 Random Matrix Theory (RMT) has been applied in surprisingly 
many fields of physics and mathematics. We have proposed in
a recent letter \cite{hm4} its application to transfer matrices
of lattice models in classical statistical mechanics.
In a preceding paper \cite{hm5},
referred to hereafter as Paper I, we have given the details of this RMT
analysis applied to the general eight-vertex model.
We have numerically shown that the  integrability of the model can be
seen on the statistical properties of the entire spectrum of
transfer matrices of vertex models. Using this as a criterion
for integrability, we have located all the known integrable
varieties in the parameter space. 
In this paper we continue this work with the study of
spin models. 
Many aspects have already been presented in paper I,
so we will recall below only the basic ideas of the RMT analysis with
emphasis on the points which are specific to spin models.

An important area of application of RMT is the characterization
of chaos~\cite{Gut90,Haa91}. One can describe the fluctuations of
energy spectra of chaotic systems with some ensembles of RMT,
while the spectra of regular systems show the characteristics
of independent numbers (Poissonian ensemble).
In classical (Hamiltonian) mechanics,
the notions of regular and chaotic dynamics coincide with
the notions of integrability and non-integrability.
But in quantum mechanics the notions of chaos and integrability are
less precise; one nowadays adopts the criterion of RMT as a definition
of quantum chaos.
For models of quantum statistical mechanics one can adopt
a definition of integrability related to the Bethe ansatz:
an integrable system is a system for which a complete set of eigenstates
having the Bethe ansatz form  
exists. For a classical statistical mechanics model, 
the notion of integrability
is generally related to the Yang--Baxter equations. For example,
solving the Yang--Baxter equations for the symmetric eight-vertex
model (also called the Baxter model) allows to build a
one-parameter family of commuting transfer matrices, and finally to
compute the free energy of the model \cite{Bax71a,Bax82}.
\nocite{Bax72a}
The spin version of the
Yang-Baxter equations are the star-triangle equations 
which have been introduced earlier \cite{Wan45,Bax82}.
Solving these star-triangle equations
also allows to construct an infinite family of commuting
transfer matrices.
We note that a vertex model can be turned into a spin
model with many-spin interactions (IRF model) (see for example \cite{LiWu72}).
In this paper, we treat spin models including only two-spin 
interactions, possibly coupled to an external field,
and by integrable we mean
Yang-Baxter integrable as well as star-triangle integrable.

The spectrum of an integrable system, after a suitable
treatment, has been shown to have many properties of a set of 
random {\em independent} numbers (Poissonian behavior), while
the spectrum of a chaotic system is described quite 
accurately by the spectrum of
matrices of statistical ensembles. 
The choice of the proper ensemble
depends on very general symmetry properties of the model
under consideration. For a time reversal symmetric model
this ensemble is the Gaussian Orthogonal Ensemble (GOE) \cite{Meh91,Haa91}.
This classification scheme has been applied succesfully to
 quantum spin models in one dimension \cite{PoZiBeMiMo93,HsAdA93}
and on two-dimensional lattices \cite{MoPoBeSi93,vEGa94,BrAdA96,BrAdA97}.

For a classical lattice spin model, the energy spectrum is usually very simple.
(For an Ising model this spectrum contains
all the possible numbers of violated bonds
i.e.\ the set of integers $[0,N_b]$ where $N_b$ is the number of bonds;
the physical properties of the model are contained in the degeneracies,
its statistical properties have been studied in Ref.~\cite{vEGa94}.)
Therefore the analysis has to be performed on another quantity. 
The transfer matrix is such an operator (related to the Hamiltonian) which
describes completely the thermodynamic properties of a system
including the size effects \cite{Transfer},
and we will  perform
the RMT analysis on its spectrum.
It has already been shown in paper I and in
\cite{hm4} that the notion of Yang--Baxter integrability
coincides with a Poissonian spectrum, and that nonintegrable
spectra are described by GOE matrix spectra. 
To  perform this RMT analysis in its usual form, one needs to deal
with a real spectrum. 
For spin models and when the interactions are nonchiral (symmetric),
this can be achieved using the so-called row-to-row
(or layer-to-layer in higher dimension) transfer matrix as
explained below. 
We also have studied the spectra in the nonphysical
regime where the Boltzmann weights are not positive. Indeed, most of the
analytical results concern  varieties in the entire parameter space
including the region where the Boltzmann weights are negative.
In this region, even row-to-row transfer matrices can have complex eigenvalues.
The necessary changes in the analysis will be briefly
discussed.

The plan of this paper is the following.
In Sec.~\ref{s:numeric} we recall the numerical
methods of the RMT analysis with a special emphasis on the
specificity of the spin models we investigate.
In Sec.~\ref{s:results} we present the numerical
results, succesively of the
two-dimensional Ising model in absence of magnetic field
(a paradigm of an integrable system!), the same two-dimensional
Ising model in a magnetic field, the three-dimensional
ising model (a paradigm of nonintegrable system!), and
the three-state Potts model. The three-state Potts model
provides an example of a point which is integrable and
critical at the same time.
By contrast, the three-dimensional Ising model provides 
a nonintegrable, but critical, point.
Finally, we investigate a nonphysical self-dual point of the
three-state Potts model.
This is a first attempt to study complex spectra in the context
of this statistical analysis of transfer matrices.
We conclude in Sec.~\ref{s:conclusion} with a discussion.


\section{Numerical Methods of RMT in the Context of Spin Models}
\label{s:numeric}

The machinery of Random Matrix Theory has been explained
in details in Paper I.
It consists in five distinct steps:
\begin{itemize}
\item
the first one is to choose a representation basis for the
operator and to construct the corresponding matrix;
\item
the second
step consists in finding the parameter-independent stable subspaces
and find the matrices representing the operator in each of these subspaces;
\item
the third step is to diagonalize each matrix;
\item
the fourth step is to `unfold' each spectrum;
\item
 and in the fifth step all the spectral
quantities are computed.
\end{itemize}
In this section, we briefly discuss these five points
 and give the details which are specific to the spin models
studied in this paper.

(i) We use the transfer matrix formalism, where the lattice is built up adding
identical ``generating sub-lattices''. These generating sublattices can
be rows for two-dimensional models or rectangular layers for
three-dimensional lattices.
For spin models, it is well known that this transfer matrix can be
factorized as:
\begin{equation} \label{e:transfer}
T(K_1,K_2,\dots,K_n) =  
	V^{1/2}(K_2,\dots,K_n) H(K_1) V^{1/2}(K_2,\dots,K_n) \;,
\end{equation}
where $H$ contains the interactions between two generating sub-lattices
and $V$ contains the interactions inside a generating sub-lattice; $V$
is usually a diagonal matrix. 
$K_i$ are the coupling constants in the different directions,
we assign the direction 1 to be the direction in which the lattice grows.
For each case we study in this paper, the precise form
of $V$ and $H$ are given in the corresponding section.
In the form (\ref{e:transfer})  the transfer matrix is a symmetric matrix
if the interaction between the two layers and thus the matrix $H$ is symmetric.
If the Boltzmann weights are real positive, 
the entries of the transfer matrix are also real positive.
In this case of a real and symmetric (transfer) matrix
the spectrum is real so that the methods for the statistical analysis
presented in paper I apply.

(ii) To find the parameter independent subspaces one needs to know
the symmetries of the generating sub-lattices.
For the transfer matrix of a $d$-dimensional hyper-cubic
lattice these generating sub-lattices are $(d-1)$-dimensional lattices.
For the square lattice, the generating sub-lattice is a periodic
chain. As explained
in Paper I the symmetry group is the dihedral group ${\cal D}_N$.
For a three-dimensional cubic lattice we use a square lattice
as the generating sub-lattice. For an isotropic square lattice
the automorphy group has been detailed in \cite{BrAdA97} (with
emphasis on the atypical case $N=4$). It is a large group which
leads to consequent size reduction (see Sec.~\ref{ss:Ising3d}).
In addition to these space symmetries there is a `color symmetry'
when no field is applied. This symmetry reflects the fact that the
Hamiltonian is invariant under a permutation of the possible values
of the spin variables
(i.e.~spin reversal symmetry for the Ising model). For
a $q$-state model, this is a $S_q$ symmetry; in this paper we work
with $q=2$ and $q=3$.
This `color-symmetry' commutes with the space symmetries.

(iii) The diagonalization of the blocks is done numerically using
standard procedures  of the LAPACK library.
We note that the block-diagonalization of step (ii) requires more
numerical effort (CPU time) than the diagonalization of one block.

(iv) The unfolding procedure is the same as the procedure detailed
in paper I. It produces the unfolded eigenvalues $\epsilon$
from the raw eigenvalues $\lambda$.
The spectra of spin and vertex models are 
quite similar. However,
we also have analyzed {\em complex} spectra. In that case 
the eigenvalues are seen as points in the plane (and
not on a line) the local density of which has to be made constant.
To unfold complex spectra we
follow the procedure described in Chap.~8.6 of Ref.~\cite{Haa91}.

(v) The spectral analysis
is performed on the same quantities as in paper I.
These are the level spacing distribution $P(s)$ of the
differences between two consecutive
unfolded eigenvalues $s_i=\epsilon_{i+1}-\epsilon_{i}$
(for a nonintegrable model,
the eigenvalue spacing distribution is very close to the Wigner surmise
for the GOE:
\begin{equation}
P(s\equiv \lambda_{i+1}-\lambda_i) = \frac{\pi}{2} s \exp( -\pi s^2 / 4) \;,
\end{equation}
in contrast, with the  exponential $P(s)=e^{-s}$ for a set of
independent eigenvalues for an integrable model);
the spectral rigidity:
\begin{equation}
\Delta_3(L) = 
\left\langle \frac{1}{L} \min_{a,b}
\int_{\alpha-L/2}^{\alpha+L/2}
{\left(N_u(\epsilon)-a \epsilon -b\right)^2 d\epsilon} \right\rangle_\alpha \;,
\end{equation}
where $N_u(\varepsilon) \equiv \sum_i \theta(\varepsilon - \varepsilon_i)$
is the integrated density of unfolded eigenvalues and
$\langle\dots\rangle_\alpha$ denotes an average over $\alpha$;
another quantity of interest 
is the number variance $\Sigma^2(L)$ defined as the
variance of the number of unfolded eigenvalues
in an interval of length $L$:
\begin{equation} \label{Sigma2def}
\Sigma^2(L) = \left\langle
\left[ N_u(\varepsilon+\frac{L}{2}) - 
       N_u(\varepsilon-\frac{L}{2}) - L \right]^2
\right\rangle_{\varepsilon},
\end{equation}
where the brackets denote an averaging over $\varepsilon$.
The expected behavior in the limiting cases of  independent numbers
and of GOE spectra has been recalled in paper I; for more details
see for example Refs.~\cite{Meh91,Haa91}.

We also recall the parametrized probability distribution we use
to interpolate between the Poisson law and the Wigner law:
\cite[ch.\ 16.8]{Meh91}:
\begin{equation}
\label{e:beta}
P_\beta(s) = c(1+\beta)\, s^\beta\, \exp\left(-c s^{\beta+1}\right)
\end{equation}
with 
$c=\left[\Gamma\left({\beta+2\over\beta+1}\right)\right]^{1+\beta}$.
The interpolation parameter $\beta$ proved itself to be a useful
indicator for the localization of integrable varieties \cite{hm5}.
There should be no confusion of this parameter $\beta$ 
with the inverse temperature $1/k_BT$.


\section{Results of the RMT Analysis}
\label{s:results}
\subsection{Two-Dimensional Ising Model without Magnetic Field}
\label{ss:Isi2dsanschamp}

We start our analysis with the two-dimensional Ising model,
which is well known to be integrable in the absence of a magnetic 
field \cite{Ons44}.
To work with symmetric matrices we use the row-to-row 
transfer matrix. For a rectangular $N \times M$ lattice with periodic
boundary conditions the (reduced) Hamiltonian reads:
\begin{equation}
\beta {\cal H}^0_{N,M}(K_1,K_2) = - \sum_{i=0}^{N-1}{
	\sum_{j=0}^{M-1}{\sigma_{i,j} \left( K_1 \sigma_{i,j+1}
		+ K_2\sigma_{i+1,j} \right)}}\;,
\end{equation}
where $\beta$ is the inverse temperature, and $K_1$ and $K_2$ are the
coupling constants in the two directions divided by the temperature.
The partition function is 
\begin{eqnarray}
Z_{N,M}(K_1,K_2)
  & = & \sum{\exp{ \left(\beta {\cal H}^0_{N,M}(K_1,K_2) \right)}} \nonumber\\
  & = & {\rm Tr} {\cal T}_N^M(K_1,K_2) \;,
\end{eqnarray}
where the row-to-row transfer matrix ${\cal T}_N(K_1,K_2)$ is given by:
\begin{equation}
{\cal T}_N(K_1,K_2) = V(K_2) H(K_1) \;,
\end{equation}
The matrix $V(K_2)$ is a $2^N \times 2^N$ diagonal matrix with entries
\begin{equation}
\label{e:V}
[V(K_2)]_{\alpha,\alpha} = \prod_{i=0}^{N-1}{w_2^{\alpha_i \alpha_{i+1}}} \;,
\end{equation}
where $\alpha_i = \pm 1$ according to the $i^{\rm th}$ digit of the
binary representation of $\alpha$ and $w_2 = e^{K_2}$.
The matrix $H(K_1)$ is a symmetric
matrix
\begin{equation}
\label{e:H}
H(K_1) = h(K_1)^{\otimes N} \;, \quad
h(K_1) = \left(
	\begin{array} {cc}
	w_1 & w_1^{-1} \\ 
	w_1^{-1}& w_1
	\end{array} 
\right) \;,
\end{equation}
with  $w_1 = e^{K_1}$.
Note that $K_1$ and $K_2$ do not play the same role here.
If all the Boltzmann weights are real and positive, and using
the circular property of the trace, ${\cal T}_N(K_1,K_2)$ can be
replaced by the similar matrix 
\begin{equation}
\label{e:ising2d}
T_N(K_1,K_2) = V^{1/2}(K_2) H(K_1) V^{1/2}(K_2) \;.
\end{equation}
It is clear that $T_N(K_1,K_2)$ is symmetric and therefore has a real
spectrum. This spectrum has been completely worked out for even
size $N$ in \cite{ScMaLi64}. From the explicit form of
the eigenvalues it is easy to check that the entire spectrum is invariant
by negating the Boltzmann weights $w_i$ and we can restrict ourselves to the 
physical case $w_i >0$.

In Ref.~\cite{ScMaLi64} it is shown that the problem of
diagonalizing $T_N(K_1,K_2)$ can be turned into a problem
of free fermions. Namely one has:
\begin{equation}
T_N(K_1,K_2) = (2 \sinh{2 K_1})^{N/2} 
   \exp{ \left(-\sum_{q=0}^{N-1}
                    {\epsilon_q \left(\xi_q^+\xi_q -{1\over 2}\right)}
         \right) } \;,
\end{equation}
where $\xi_q$ are fermionic operators. The dispersion relation
$\epsilon_q(K_1,K_2)$ is a cosine and its detailed form depends
on the parity of the number of $\xi$-particles. Therefore, fixing
all the``quantum number'' $q_i$, where $q_i$ is the momentum of the
$i^{\rm th}$ $\xi$-particle, {\em completely} determines an eigenvalue and
an eigenstate of the transfer matrix. However, instead of using the
complete set of quantum numbers, we take only into account the 
quantum numbers which correspond to parameter-independent symmetries.
As mentioned in Sec.~\ref{s:numeric},
the space symmetries of $T_N(K_1,K_2)$
form a group isomorphic to 
${\cal D}_N = Z_N \,\rule[0.02ex]{0.06ex}{1.15ex}\!\mbox{$\times$} Z_2$,
and in the 
absence of a magnetic field the spin reversal symmetry induces
an extra $Z_2$ symmetry which commutes with ${\cal D}_N$. Using the 
projectors independent of $K_1$ and $K_2$ onto the corresponding
invariant subspaces, the transfer matrix is block diagonalized. The
dimensions of the blocks are given in Table \ref{t:Ising2d}.
We note that using these projectors leaves a few degeneracies inside
some blocks. We also note that some eigenvalues are independent
of the value of $K_2$. In Appendix~\ref{a:appendix} all
these $2^{N/2+1}$ $K_2$-independent eigenvalues occuring only
when the size $N$ is a multiple of four, are
determined analytically.

\begin{figure}
\psfig{file=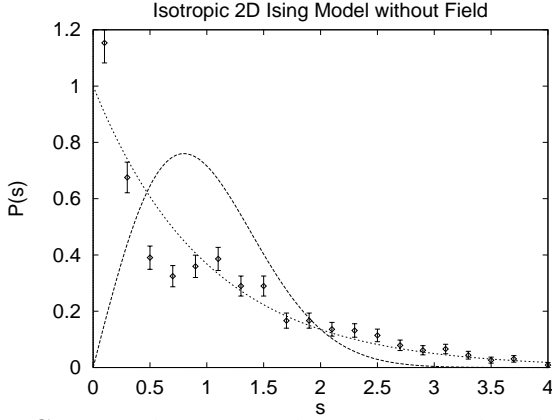,width=\hsize}
\caption{The eigenvalue spacing distribution for the row-to-row
transfer matrix of the isotropic two-dimensional Ising model.
The linear size is $N=16$ and the Boltzmann weight is $w_1=w_2=1.4$.
The spacings are taken in the largest representation.
The average is over 1400 observations.}
\label{f:fig1}
\end{figure}

Having discarded the degenerate states and the $K_2$-independent
states in each representation one can perform the RMT analysis.
Fig.~\ref{f:fig1} shows the level spacing distribution for $w_1=w_2=1.4$
of the  representation labelled $R=4, C=0$ of $N=16$
(see Table~\ref{t:Ising2d}).
One finds roughly an exponential distribution which is expected for an
integrable model.
Our explanation for the deviation from the exponential is the following:
for an integrable system we do not sort the states according to all
their ``quantum numbers'' (doing that will leave us with blocks of
size 1). We therefore treat together eigenvalues belonging to states
having different symmetries, but these states are only approximately
independent.
We also have performed the same numerical analysis on {\em all}
eigenvalues computed with formula (\ref{e:Lambda}) and the result is even
worse.
This result is characteristic of the free-fermion nature of the
problem (note that the parameter $\beta$ has been found always negative
on the free-fermion variety of the eight-vertex model \cite{hm5}).
The same form of the spacing distribution is found for any value of the
Boltzmann weights, even for negative Boltzmann weights.
However, if one of the Boltzmann weight is `too large' or `too small'
some entries of the transfer matrix become huge and this leads to 
numerical instabilities in the diagonalization. On the other hand,
if the Boltzmann weights are `too close' to unity (decoupling limit)
many eigenvalues are almost-degenerate which leads to difficulties
in the  unfolding procedure. We note finally that the critical point does not
manifest itself in any manner on the spacing distribution
nor on the spectral rigidity $\Delta_3$.

\subsection{Two-Dimensional Ising Model in a Magnetic Field}
\label{ss:Isi2davecchamp}

We now investigate the case where a magnetic field is turned on.
The Hamiltonian becomes:
\begin{equation}
\beta {\cal H}_{N,M}(K_1,K_2) = \beta {\cal H}^0_{N,M}(K_1,K_2) - 
			K \sum_{i=0}^{N-1}{\sum_{j=0}^{M-1}{\sigma_{i,j}}}\;,
\end{equation}
$K$ being the field times the inverse temperature,
and the transfer matrix reads
\begin{equation}
T_N(K_1,K_2,K) = [V(K_2)B(K)]^{1/2} H(K_1) [V(K_2)B(K)]^{1/2}\;,
\end{equation}
where $H$ and $V$ are the same matrices as
defined by Eq. (\ref{e:H}) and (\ref{e:V})
and $B(K)$ is a diagonal matrix with entries
\begin{equation}
[B(K)]_{\alpha,\alpha} = \prod_{i=0}^{N-1}{w^{\alpha_i}}
\end{equation}
where $\alpha_i=\pm 1$ according
to the $i^{\rm th}$ bit of the binary representation of $\alpha$,
and $w=e^K$.
The spin reversal symmetry does not hold any more.

\begin{figure}
\psfig{file=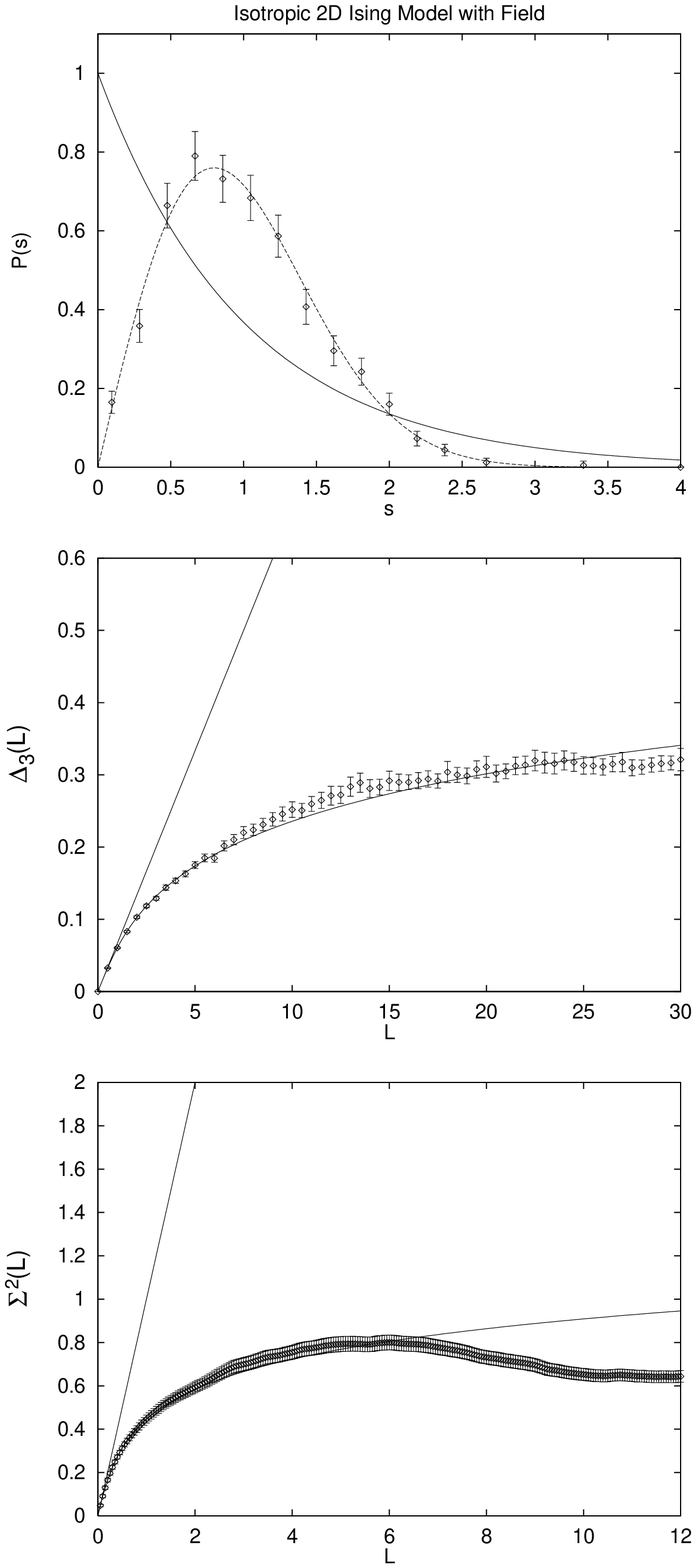,width=\hsize}
\caption{The eigenvalue spacing distribution $P(s)$, the rigidity $\Delta_3$
and the
number variance $\Sigma^2$ for the row-to-row
transfer matrix of the isotropic two-dimensional Ising model
in a field.
The linear size is $N=14$, the temperature is $T=1.4$ and the magnetic
field is $H=0.8$.
The spacings are taken in the representation labelled $R=4$.
The average is over 1100 observations.}
\label{f:fig2}
\end{figure}

\begin{figure}
\psfig{file=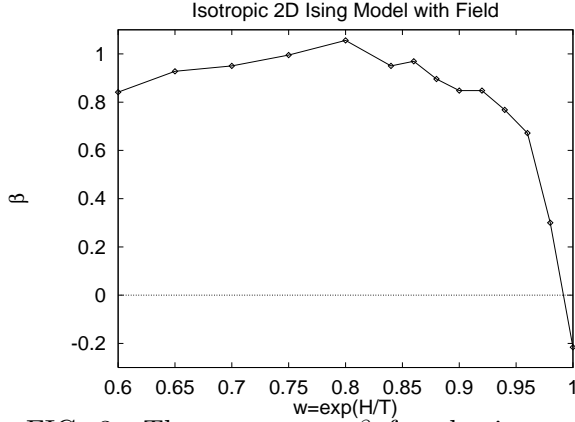,width=\hsize}
\caption{The parameter $\beta$ for the isotropic square lattice
Ising model as a function of the Boltzmann weight $w=\exp(H/T)$
associated with the field. The other Boltzmann weights are kept
constant $w_1=w_2=1.6$ which determines the temperature as $T\approx 2.1$.
Average over the representations $R=0$ and $R=4$ for $N=14$
(approx 1700 spacings).}
\label{f:figx}
\end{figure}

It is known that this model is not Yang--Baxter integrable and its
partition function has not yet been summed up. We present in 
Fig.~\ref{f:fig2} the spacing distribution and the spectral rigidity
for a typical large representation (labelled $R=4$ in Table \ref{t:Ising2d})
for $L=14$. The temperature is $T=1.4$ and the magnetic field
is $H= 0.8$. The statistics is taken over 1100 spacings.
The spacing distribution clearly coincides
with the Wigner Surmise. The agreement of the spectral rigidity
$\Delta_3$ and the number variance $\Sigma^2$ 
with same quantities computed for the GOE matrices is
surprisingly good: it holds up to a value of $L=25$ for the
spectral rigidity and only up to $L=7$ for the number variance,
a value much larger than for other models of statistical
mechanics \cite{hm5}. To appreciate how the magnetic field influences
the spacing distribution we have plotted in Fig.~\ref{f:figx}
the best fitted value
$\beta$ of the Brody distribution $P_\beta$ Eq.~(\ref{e:beta})
as a function of the Boltzmann weight associated to the field.
The behavior of $\beta$ is unambiguous: the magnetic field induces
a Wigner type distribution of the spacing distribution. The drop of
$\beta$ is sharp as the magnetic field goes to zero.
On Fig.~\ref{f:figx} the negative value of the parameter $\beta$ in zero field 
is due to the free-fermion nature of the two dimensional
Ising model in absence of magnetic field.
We interpret the small width of the peak as a size effect, and we claim
that in the thermodynamic limit $\beta$ is strictly zero only
for a zero magnetic field. This underlines the very singular nature
of integrability. Probing the value of $\beta$ in the
physical region of the parameter space, we did not have find other
integrable points than the points where the magnetic field is zero.

\subsection{Three-Dimensional Ising Model}
\label{ss:Ising3d}
We now investigate another archetypal case of non-integrability
in statistical mechanics: the three-dimensional Ising model.
This model can still be mapped onto a fermion problem
like in two dimensions,
but in three dimensions the fermions are correlated \cite{Itz82b}
and the partition function cannot be summed up in a closed form.

We build the lattice by adding {\em square isotropic} layers.
Be $K_2$ the interaction in the two directions inside the layers and
$K_1$ the interaction in the direction perpendicular to the layers
(i.e.\ the cubic lattice is built in the direction of $K_1$).
We again have 
$T_N(K_1,K_2) = V^{1/2}(K_2) H(K_1) V^{1/2}(K_2)$. 
The matrix $V(K_2)$ is a $2^{N^2} \times 2^{N^2}$ diagonal 
matrix with entries
\begin{equation}
[V(K_2)]_{\alpha,\alpha} = \prod_{i=0}^{N-1}\prod_{j=0}^{N-1}
{w_2^{\alpha_{i,j} (\alpha_{i,j+1}+\alpha_{i+1,j})}}
\end{equation}
where $\alpha_{i,j} = \pm 1$ (the sums in the indices are
taken modulo $N$).
The matrix $H(K_1)$ is a symmetric
matrix with entries
\begin{equation}
[H(K_1)]_{\alpha,\beta} = 
    \prod_{i,j=0}^{N-1}{w_1^{\alpha_{i,j} \beta_{i,j}}} \;.
\end{equation}
We again need to project the transfer matrix into the 
parameter independent invariant subspaces. Without magnetic field
we still have the spin reversal symmetry, but the space group is
more involved than for the two-dimensional model, since the automorphy
group of a square lattice is larger than the automorphy group of
a ring (see Sec.~\ref{s:numeric}). 
Using rectangular $N \times N'$ layers would have led to 
much simpler calculations
since the symmetry group would have been simply 
${\cal D}_N \otimes {\cal D}_{N'}$.
However, the size reduction of the matrix would have been less
and consequently also the numerically accesible lattice sizes. It is worth
noting at this point that the size effects 
are not controlled by the dimension
of the subblocks of the transfer matrix, but by the size of the lattice.
Moreover, we have studied
the particular case $N=4$ for which the 
generic symmetry group of the isotropic square lattice $C_{4v}$
has to be replaced by a larger group
(for the technical details see \cite{BrAdA97}).
In contrast with the two-dimensional case, projecting onto invariant
subspaces lifts {\em all} degeneracies within each subblock 
(note that using the generic $C_{4v}$ group would have left
degeneracies inside some blocks).
Table \ref{t:Ising3d} shows the size of the invariant subspaces.

\begin{figure}
\psfig{file=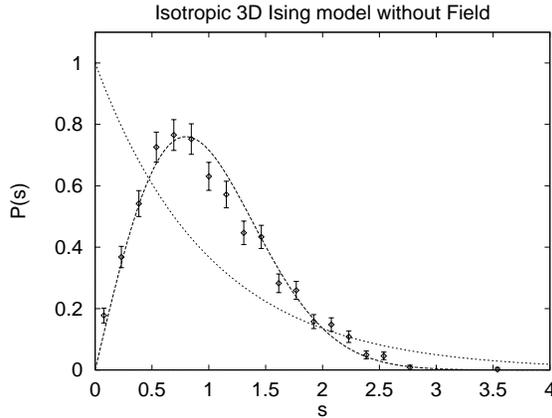,width=\hsize}
\caption{The eigenvalue spacing distribution for the row-to-row
transfer matrix of the isotropic three-dimensional Ising model
at the critical point.
The linear size is $N=4$.
The spacings are averaged over the representations labelled 17 and 18.
The average is over 2000 observations.
}
\label{f:fig3}
\end{figure}

The results of the RMT analysis are presented in Fig.~\ref{f:fig3}.
The eigenvalue spacing distribution for the
representations labelled $R=17$ and $R=18$
in Table \ref{t:Ising3d} are averaged together. After
discarding some eigenvalues close to the edge of the spectrum
we are left with 2000 spacings.
We observe that the level spacing distribution is 
very close to a Wigner
distribution. Also the spectral rigidity and the number variance
are in agreement with the corresponding quantities of GOE matrices.
We have paid special attention to the critical
point, since it is an example of a nonintegrable, nevertheless critical,
point. No special behavior is found.
This strongly suggests that the statistical properties of the
eigenvalues of transfer matrices are governed by the status of the model
with respect to integrability rather than to criticality.

\begin{figure}
\psfig{file=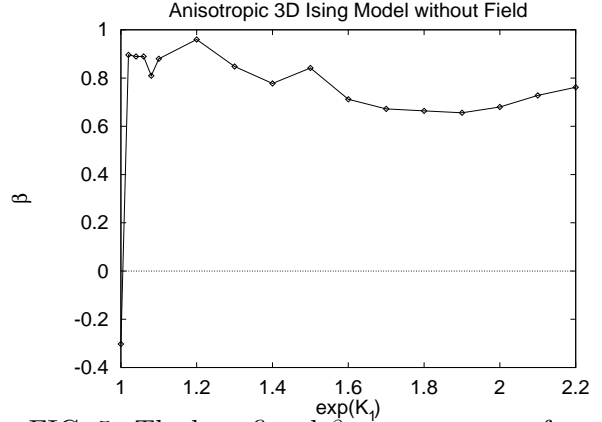,width=\hsize}
\caption{The best fitted $\beta$ parameter as a fucntion of $e^{K_1}$
for the anisotropic three-dimensional Ising model.
The average runs over 2000 spacings. The point $e^{K_1}=1$
corresponds to a two-dimensional model.
}
\label{f:fig4}
\end{figure}

To go from a two-dimensional model to a three-dimensional model
we have recorded the value of $\beta$ when the interaction $w_1$
in the third dimension is turned on continuously. Fig.~\ref{f:fig4} 
summarizes the results: a vanishingly small coupling in the third direction
induces level repulsion. It is remarkable that with such a small size
$N=4$ we already have a very abrupt variation of $\beta$. This
once again stresses the singular nature of integrability.

\subsection{Three-State Potts Model on a Square Lattice.}

We now turn to the case of the Potts model (see Ref.~\cite{Wu82}
and reference therein).
This spin model is a generalization of the Ising model where the
spins can take more than two values.
The Hamiltonian is:
\begin{equation}
\beta {\cal H}^0_{N,M}(K_1,K_2) = - \sum_{i=0}^{N-1} \sum_{j=0}^{M-1}
	\bigl(K_1 \delta(\sigma_{i,j},\sigma_{i,j+1})
	      + K_2 \delta(\sigma_{i,j},\sigma_{i+1,j}) \bigr) \;,
\end{equation}
where $\delta$ is a Kronecker symbol, $\beta$ is the inverse
temperature, $K_1$ and $K_2$ are the coupling constants divided
by the temperature and $\sigma \in Z_q$ can take $q$ values.
 It has been investigated by
many authors but its full solution is still a challenge.
Using a duality relation one can localize a phase transition
at the temperature $T_c=1/\ln(1+\sqrt{q})$
for any number $q$ of states \cite{Pot52}.
The model can be mapped onto a staggered six-vertex model,
the parameters of which depend on $K_1$, $K_2$ and $q$ \cite{Bax82}.
This six-vertex model is in general not integrable since there are
two different sets of Boltzmann weights, one for each sublattice
of the square lattice. However, for a special combination of the parameters 
the two sets of Boltzmann weights are the same and, consequently, the
partition function of the  model can be calculated.
This line where the partition function can be calculated
turns out to be the critical line. 
Here, in contrast with the three-dimensional 
Ising model, the two notions of integrability
and criticality coincide.

\begin{figure}
\psfig{file=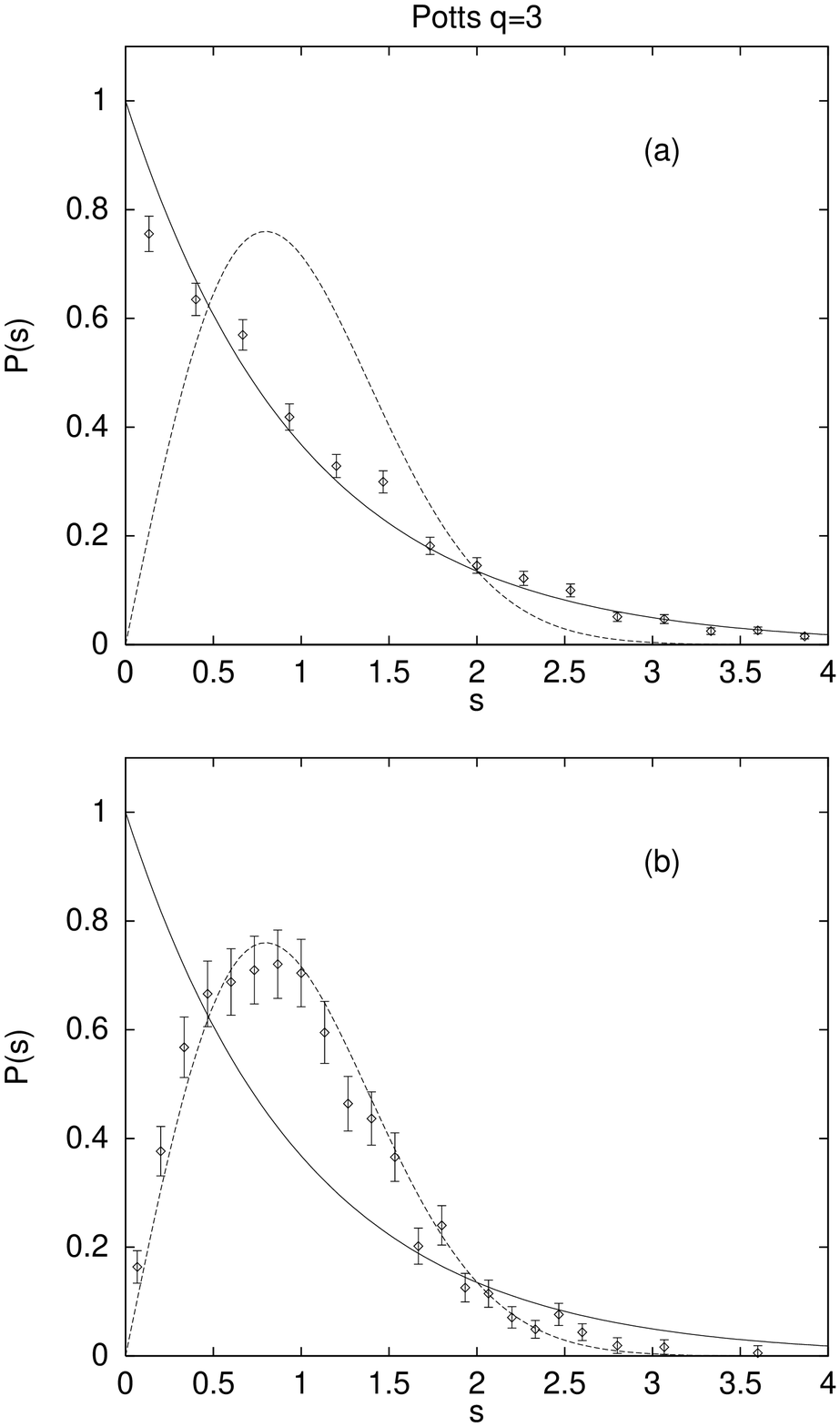,width=\hsize}
\caption{
Eigenvalue spacing distribution for the
three-state Potts model (a) precisely at 
the critical point $w_1=w_2=1+\protect\sqrt{3}$ 
and (b) far from the critical point at $w_1=w_2=1.4$. The data are obtained
for $N=11$ and the number of spacings is $2500$ (a) and 1400 (b).
}
\label{f:fig5}
\end{figure}

\begin{figure}
\psfig{file=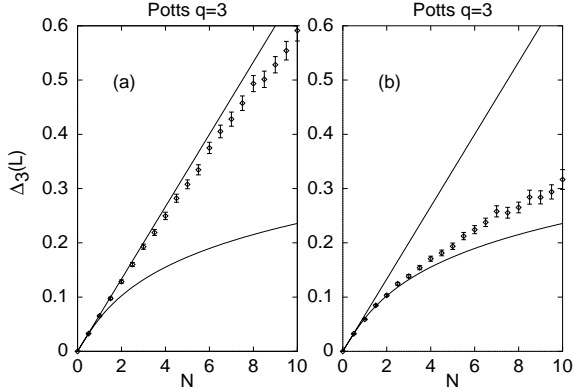,width=\hsize}
\caption{The rigidity $\Delta_3(L)$ for the same parameters as
in figure \protect\ref{f:fig5}.}
\label{f:fig6}
\end{figure}

\begin{figure}
\psfig{file=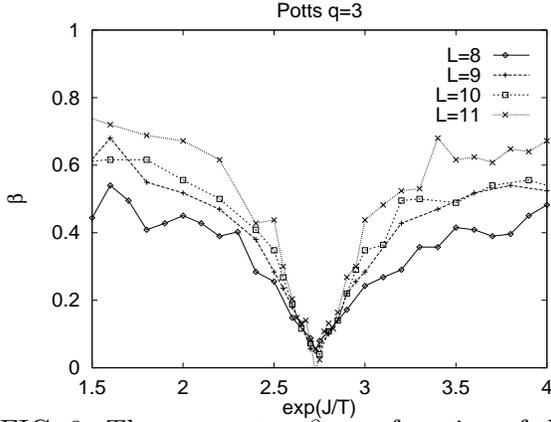,width=\hsize}
\caption{The parameter $\beta$ as a function of the
temperature for a three-state Potts model for different lattice
size. The minimum is at the critical point which is integrable.
}
\label{f:fig7}
\end{figure}

We have numerically investigated the case of the three-state
Potts model. For $q=3$ this model presents a second order phase
transition which is not of the same universality
class as the transition of the Ising model. For the isotropic case
the transition is given by $(e^K - 1)^2 = q=3$.
It would have been interesting to study higher values of
$q$ for which the transition becomes first order,
but the size of the transfer matrix is an exponential function
of the number of states $q$, and values of $q$ larger than 3
lead to extremely large matrices even for a small lattice size.
The transfer matrix $T$ has the same form (Eq.~(\ref{e:ising2d})) 
as for the two-dimensional Ising model in absence of magnetic field.
The matrix $V(K_2)$ is a $3^N \times 3^N$ diagonal matrix with entries
\begin{equation}
\label{e:VPotts}
[V(K_2)]_{i,j} = \prod_{i=0}^{N-1}{w_2^{\delta(\alpha_i, \alpha_{i+1})}} \;,
\end{equation}
where $\alpha_i$ = 0, 1 or 2 is the value of the $i^{\rm th}$ 
spins in the spin configuration labelled by $\alpha$.
The matrix $H(K_1)$ is a symmetric
matrix:
\begin{equation}
H(K_1) = h(K_1)^{\otimes N} \;, \quad
h(K_1) = \left(
	\begin{array} {ccc}
	w_1 & 1   & 1 \\ 
	1   & w_1 & 1 \\
	1   & 1   & w_1
	\end{array} 
\right) \;.
\end{equation}
The space symmetry group is ${\cal D}_N$ as in the case of
the Ising model, but the color symmetry group is $S_3$ rather
than $S_2$ for the Ising model. The size reduction is better
in the Potts case but far less than the exponential increase
due to the three states. Tab. \ref{t:Potts} gives the size of the blocks
in the different invariant subspaces.
On Fig.~\ref{f:fig5} we present two typical level
spacing distributions. The upper one (a) is obtained at the critical value
of the Boltzmann weight $w^*=1 + \sqrt{3}$ and the other one is obtained
at a different value $w=1.4$ far from the transition.
It is obvious that at $w^*$ the distribution is very close to
an exponential while for $w \neq w^*$ this distribution is
close to the Wigner surmise.  Fig.~\ref{f:fig6}
shows the rigidity $\Delta_3$ for the same 
values of the Boltzmann weights.
We observe the same coincidence with the theoretical
behavior expected for independent numbers 
and spectra of GOE matrices. 
The agreement with
the GOE behavior extend up to a value $L \simeq 4$.
This value is much less than for the two-dimensional Ising
model in a field, but are comparable with values obtained
in some quantum models \cite{BrAdA96}.
We present on Fig.~\ref{f:fig7} the behavior of
the best fitted value $\beta$ of the parametrized distribution
as a function of temperature. Several sizes $N=$ 8, 9, 10 and 11
are plotted. We first note that the spacing distribution has properly
`detected' the integrable point $w^*$, which corresponds 
precisely to the minimum in the $\beta(w)$ curve. We also observe
size effects: for larger size the downward peak is sharper than
for smaller size. This suggests that in the thermodynamic limit
the spacing distribution is a Poisson law {\em only} at the
critical temperature.

We have again found that integrability
leads to independent eigenvalues, whereas non-integrability
leads to eigenvalue repulsion properly described by
the spectral statistical properties of the GOE: the
analysis of the statistical properties of the spectrum
of the transfer matrix can be used to find integrable points.

\subsection{Three-State Potts Model for Negative Boltzmann Weights}

We now look at the isotropic three-state Potts model with $w<0$.
The transfer matrix is still symmetric, but some
entries  become imaginary, since we have
half-integer powers of negative numbers 
(see Eqs.~(\ref{e:VPotts}) and (\ref{e:ising2d})).
We have numerically found
that the spectrum is complex only when $-2<w<0$.
So when $w<-2$ one can apply the standard RMT analysis.
We have found a GOE spacing distribution as expected.
When $-2<w<0$ the spectrum contains 
mostly complex conjugate eigenvalues.
By spacing we now mean shortest euclidian distance between
eigenvalues in the complex plane, and we use another
unfolding procedure \cite[Chap. 8.6]{Haa91}.
We want to know if the unfolded eigenvalues are `independent'
or if they repel each other. The spacing distribution 
of independent points in $d$ dimensions is easy to evaluate:
for $N$ points  taken randomly on a $d$-dimensional 
hyper-sphere of radius $N^{1/d}$, the probability that the 
distance between a point and its closest neighbor
is $s$, is the probability
that exactly one point is found at the distance between $s$ and $s+ds$,
and that the other $N-2$ points are farther away than $s$:
\begin{equation}
P_N(s)  d \! s  \propto s^{d-1} 
\left[ 1-\left(\frac{s}{N^{1/d}}\right)^d 
\right]^{(N-2)} d \! s  \;.
\end{equation}
Taking the limit $N \rightarrow \infty$ one gets 
$P(s) \propto s^{d-1} \exp{(-C s^d)}$ ($C$ is some constant).
For $d=2$ one recovers precisely the Wigner law which is a well known
distribution in mathematical statistics.
In other words a Wigner law for a one-dimensional set of points
means repulsion between these points, whereas 
for points in the plane the same Wigner law means independence.
The probability that random points in the plane are close to each other
is already small and a supplementary repulsion due to correlations of
non-integrability will have less influence than in the case of a
real spectrum.
For  nonsymmetric random matrices the joint probability distribution
of the eigenvalues has been studied \cite{LeSo91}.
Eigenvalue repulsion is still present, but a closed expression
for the degree of repulsion is not known.

\begin{figure}
\psfig{file=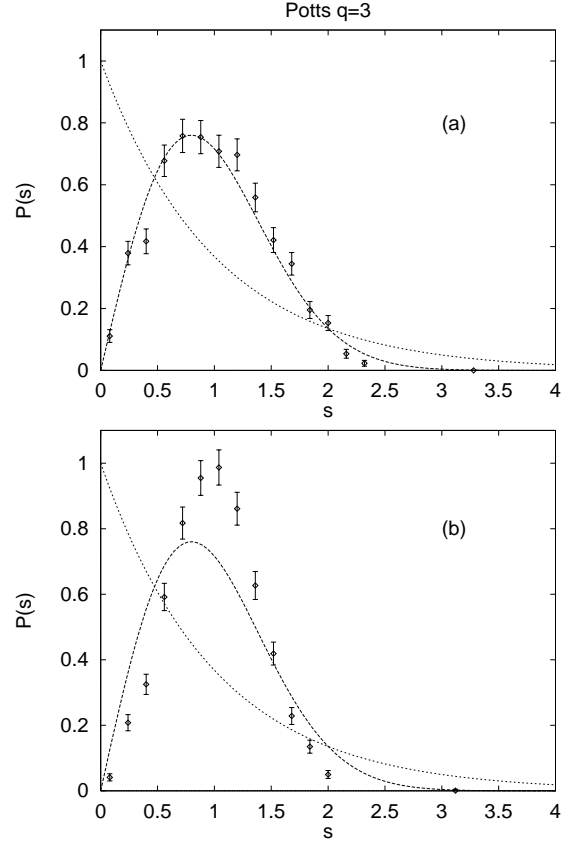,width=\hsize}
\caption{Two closest-distance distributions for
the isotropic three-state Potts model
with negative Boltzmann weight:
(a) for the self-dual Boltzman weight 
$w_1=w_2=w_{SD}=1-\protect\sqrt{3}$ and (b) for a
different value $w_1=w_2=-1.5$. The average runs over
about $2000$ spacings.}
\label{f:fig8}
\end{figure}

On Fig.~\ref{f:fig8}(b) we present a distribution of eigenvalue
spacings for $w=-1.5$: the repulsion
is clearly seen, since $P(s)$ near the origin is smaller than for
the Wigner law. We note that this distribution is not
close to the eigenvalue spacing of the spectra of GUE matrices.
We also have investigated another special point
in the regime $-2<w<0$. The self-duality equation $(w_{SD}-1)^2 = q$
has a solution $w_{SD} = 1-\sqrt{3}$ which lies in this regime.
On Fig.~\ref{f:fig8}(a) we present the distribution of eigenvalues spacings
at this value $w_{SD} = 1-\sqrt{3}$. The agreement with the Wigner law,
which here means independence, is quite good. This is
expected since the point $w_{SD}$ is also integrable.

\section{Discussion and Conclusion}
\label{s:conclusion}

We have numerically shown that the eigenvalues of the transfer matrix
of an integrable spin model have many features of random independent 
numbers. 
On the contrary, the spectrum of the transfer matrix
of a nonintegrable spin model has many features in common with
the spectrum of a GOE matrix.
In particular, the spacing distribution is an exponential law for
integrable systems while it is close to the Wigner surmise
for nonintegrable systems. Quantities involving more
than two closest eigenvalues, like the rigidity, also show a GOE
behavior on a quite large scale involving up to 25 eigenvalues
in the case of the square lattice Ising model with field.
The independence of eigenvalues for integrable models
has been checked on the Ising model in two dimensions in
absence of a magnetic field, as well as on the critical point
of the three-state Potts model. We also have studied the transition point
of three-dimensional Ising model, for which we 
have found eigenvalue repulsion and a
good agreement of the spacing distribution with the Wigner surmise.
This stresses the difference between criticality and integrability:
the eigenvalue statistics is sensible to integrability and not to criticality.

Using the eigenvalue statistics as a criterion of integrablity we support
the hypothesis that the two-dimensional Ising model in a field
and the three-dimensional Ising model are not integrable.
The integrable models appear as very singular and isolated in
parameter space. This is clearly seen on 
curves showing the parameter $\beta$
as a function of a Boltzmann weight, where $\beta$ is close to
unity almost everywhere, except for the particular values where
the model is integrable. With the sizes numerically tractable
the variation of $\beta$ is abrupt, and the size behavior suggests
that in the thermodynamic limit the statistics changes discontinuously.
To clearify this point, a more detailed study of the size effects is needed.
It would also be interesting, but difficult, to study the Potts
model with a large number of states to have a first order phase transition
point which is integrable.
 
To use the criterion of eigenvalue spacing statistics with new models
(for example chiral models)
one needs to study complex spectra in many cases.
The distinction between independent eigenvalues for integrable
models and repelling eigenvalues for nonintegrable models
still holds. However, a repulsion between eigenvalues in one dimension
is much easier to quantify than in two dimensions. 
Intuitively this can be understood
since in two dimensions the eigenvalues are not restricted to a line
and have naturally more space to avoid each other.
The repulsion between eigenvalues in two dimensions has less effect than
in one dimension. We have found that for the three-state Potts model
and for a Boltzmann weight $-2 < w < 0$ the spectrum is complex
and that the eigenvalue spacing distribution 
is characteristic of eigenvalue repulsion.
However, when $w$ is close to the negative self-dual value
we recover a distribution close to the Wigner law indicating here
eigenvalue independence associated to integrability. 
The numerical difficulties arising with complex spectra are
of importance and we need to refine further the analysis,
and especially the unfolding, to study convicingly chiral models.
This is in progress.

In these two papers
we have shown numerically that the statistical properties
of transfer matrix spectra of classical statistical mechanics models
are related to the integrability of
the model, extending hereby the field of application of RMT.
This can be useful in the search for new integrable models.

\acknowledgements
We thank J.M.~Maillard for many discussions where his expertise
with integrable models has been very precious.

\appendix
\section{}
\label{a:appendix}
It has been established in \cite{ScMaLi64} that the spectrum
of the row-to-row anisotropic Ising model for an even size $N$ is
given by:
\begin{equation} \label{e:Lambda}
\Lambda_{\tau}(K_1,K_2) = \left[ 2 \sinh{(2 K_1)} \right] ^{N/2}
	\exp\left(- \frac{1}{2} \sum_{q=0}^{N-1}{\tau_q \epsilon_q}\right)
\end{equation}
where $\tau_q = \pm 1$ according to the $q^{\rm th}$ digit of the
binary representation of $\tau$ ($0 \leq \tau < 2^N$), and
for all $q$ with $0 \leq q < N$:
\begin{equation}
\label{e:app}
\cosh{\epsilon_q} = \cosh{2K_2} \cosh{2K_1^*}
 	-\sinh{2K_2} \sinh{2K_1^*} \cos{Q} \;,
\end{equation}
where $Q = (2 q + 1) \pi /N$ if there is an even number of 1 in the
base two representation of  $\tau$
and $Q = 2 q \pi /N$ otherwise.
For $\epsilon_q$, the positive root of 
Eq.~(\ref{e:app}) has to be taken except
for $Q=0$ where $\epsilon_0 = 2(K_1^* - K_2)$ and for
$Q = \pi$ where $\epsilon_{N/2} = 2(K_1^* + K_2)$. $K_1^*$ is the dual
coupling constant defined by $\tanh{K_1} = e^{-2K_1^*}$.
The derivative of $\Lambda_{\tau}(K_1,K_2)$ with respect to $K_2$ is
\begin{equation} 
\frac{\partial \Lambda_{\tau}(K_1,K_2)}{\partial K_2} =
- \frac{1}{2} \; \Lambda_{\tau}(K_1,K_2) 
\sum_{q=0}^{N-1}{ \tau_q \frac{\partial  \epsilon_q}{\partial  K_2} }
\end{equation}
with 
\begin{equation}
\frac{\partial  \epsilon_q}{\partial  K_2}
   = 2 \, \frac{\sinh{2 K_2}\cosh{2 K_1^*}
      - \cosh{2 K_2} \sinh{2 K_1^*} \cos{Q}}
            {\sinh{\epsilon_q}}
\end{equation}
It is easy to check that for an {\em even} number of particles:
\begin{equation}
\label{e:even}
\epsilon_q = \epsilon_{N-1-q} \;\;\;{\rm and}\;\;
\frac{\partial  \epsilon_q}{\partial  K_2} =
	\frac{\partial  \epsilon_{N-1-q}}{\partial  K_2} \qquad
(0 \leq q < N/2) \; ,
\end{equation}
while for an {\em odd} number of particles:
\begin{equation}
\label{e:odd}
\epsilon_q = \epsilon_{N-q}\;\;\;{\rm and}\;\;\;
\frac{\partial  \epsilon_q}{\partial  K_2} =
	\frac{\partial  \epsilon_{N-q}}{\partial  K_2}\qquad
(1 \leq q < N/2) \;,
\end{equation}
and 
\begin{eqnarray} \label{e:odd1}
\epsilon_0 = 2(K_1^* - K_2)\;,\;\; &&
\frac{\partial  \epsilon_0}{\partial  K_2} =2\;, \\
\epsilon_{N/2} = 2(K_1^* + K_2)\;,\;\; &&
\frac{\partial  \epsilon_{N/2}}{\partial  K_2} =-2 \;.
\end{eqnarray}

When the size $N$ is a multiple of 4 and when
there are $N/2$ particles, then the number of particles
is even and we must use the formula (\ref{e:even}). There are
$2^{N/2}$ choices of $\tau$ such that $\tau_q = -\tau_{N-1-q}$ 
for $0 \leq q < N/2$.
In that case one has:
\begin{equation}
\sum_{q=0}^{N-1}{ \tau_q \frac{\partial  \epsilon_q}{\partial  K_2} }
=
\sum_{q=0}^{N/2-1}{ (\tau_q -\tau_{N-1-q}) 
\frac{\partial  \epsilon_q}{\partial  K_2} }
= 0 \;.
\end{equation}
This gives $2^{N/2}$ $K_2$-independent states. 

$N$ being still a multiple of 4 and taking $N/2 \pm 1$ particles
we now must use  formul\ae\ (\ref{e:odd}) and (\ref{e:odd1}).
There are $2 \times 2^{N/2-1}$
choices of $\tau$ such that $\tau_q = -\tau_{N-q}$ for $1 \leq q < N/2$
and $\tau_0 = \tau_{N/2}$.
In that case one has 
\begin{equation}
\sum_{q=0}^{N-1}{ \tau_q \frac{\partial  \epsilon_q}{\partial  K_2} }
=
\pm (\epsilon_0 + \epsilon_{N/2})
+ \sum_{q=1}^{N/2-1}{ (\tau_q -\tau_{N-1-q})
 \frac{\partial  \epsilon_q}{\partial  K_2} }
= 0  \;.
\end{equation}

Collecting the three cases mentioned above we get $2^{N/2+1}$
$K_2$-independent states. It is then simple to verify that the 
corresponding eigenvalues are
\begin{equation}
2^N \, (\sinh{K_1})^p (\cosh{K_2})^{N-p} \;,
\end{equation}
where $p=N/2-1$, $N/2$ or $N/2+1$ is the number of particles.


\pagebreak

\begin{table}

\begin{tabular}{ccccrrr}
\multicolumn{7}{c}{Ising Model on a Periodic Square Lattice}\\
\multicolumn{7}{c}{N=14 \quad dim=16384}  \\ \hline
 R & k & $\lambda$ & $l_R$ & $a_{R,C=0}$ & $a_{R,C=1}$ & $a_{R}$ \\ \hline
0 &     0     & 1 & 1 & 362 & 325 & 687\\
1 & $ \pi   $ &$-1$ & 1 & 288 & 325 & 613\\
2 &$   0   $ &$-1$ & 1 & 234 & 261 & 495\\
3 &$ \pi   $ & 1   & 1 & 288 & 261 & 549\\
4 &$ 2 \pi/7$  & *  & 2 & 594 & 585 & 1179\\
5 &$ 4 \pi/7$  & *  & 2 & 594 & 585 & 1179\\
6 &$ 6 \pi/7$  & *  & 2 & 594 & 585 & 1179\\
7 &$ \pi/7$  & *  & 2 & 576 & 585 & 1161\\
8 &$ 3 \pi/7$  & *  & 2 & 576 & 585 & 1161\\
9 &$ 5 \pi/7$  & *  & 2 & 576 & 585 & 1161\\ \hline
\hline
\multicolumn{7}{c}{N=16 \quad dim=65536} \\ \hline
 R & k & $\lambda$ & $l_R$ & $a_{R,C=0}$ & $a_{R,C=1}$ & $a_{R}$ \\ \hline
0 &     0     & 1   & 1 & 1162 & 1088 & 2250\\
1 & $ \pi   $ &$-1$ & 1 & 1033 & 1088 & 2121\\
2 & $   0   $ &$-1$ & 1 & 906 & 960 & 1866\\
3 & $ \pi   $ & 1   & 1 & 1033 & 960 & 1993\\
4 & $ \pi/2 $ & *  & 2 & 2065 & 2048 & 4113\\
5 & $ \pi/4$  & *  & 2 & 2062 & 2048 & 4110\\
6 & $3\pi/4$  & *  & 2 & 2062 & 2048 & 4110\\
7 & $ \pi/8$  & *  & 2 & 2032 & 2048 & 4080\\
8 & $7\pi/8$  & *  & 2 & 2032 & 2048 & 4080\\
9 & $5\pi/8$  & *  & 2 & 2032 & 2048 & 4080\\
10 & $3\pi/8$ & *  & 2 & 2032 & 2048 & 4080
\end{tabular} %
\caption{Two-dimensional Ising model: the symmetries of a periodic chain 
combined with the $Z_2$ color symmetry.
The dimensions $a_R$ and degeneracies $l_R$ of the
invariant subspaces for $N=14$ and $N=16$.
$R$ is an arbitrary label of the representations of
the dihedral group, $\exp(ik)$ and $\lambda$ are the eigenvalues
of the corresponding translation and reflection operators
(* means that the corresponding representation is not
stable under the action of the reflection operator).
The column $a_{R,C=0}$ (resp.\ $a_{R,C=1}$) refers to states which are 
even  (resp.\ odd) under spin reversal.
The column  $a_{R}$ is the sum of the two preceding for the case
where a magnetic field breaks the spin reversal symmetry.
}
\label{t:Ising2d}
\end{table}

\begin{table}
\begin{tabular}{ccrrr}
\multicolumn{5}{c}{Ising Model on a Periodic Cubic Lattice}\\
\multicolumn{5}{c}{N=4 \quad dim=65536} \\ \hline
 R &  $l_R$ & $a_{R,C=0}$ & $a_{R,C=1}$ & $a_{R}$ \\ \hline
0 & 1 & 222 & 180 & 402\\
1 & 1 & 50 & 44 & 94\\
2 & 1 & 169 & 180 & 349\\
3 & 1 & 33 & 44 & 77\\
4 & 2 & 191 & 192 & 383\\
5 & 2 & 211 & 192 & 403\\
6 & 3 & 186 & 180 & 366\\
7 & 3 & 291 & 300 & 591\\
8 & 3 & 183 & 180 & 363\\
9 & 3 & 354 & 300 & 654\\
10 & 4 & 460 & 480 & 940\\
11 & 4 & 236 & 224 & 460\\
12 & 4 & 460 & 480 & 940\\
13 & 4 & 236 & 224 & 460\\
14 & 6 & 397 & 416 & 813\\
15 & 6 & 507 & 480 & 987\\
16 & 6 & 447 & 480 & 927\\
17 & 6 & 681 & 672 & 1353\\
18 & 8 & 668 & 672 & 1340\\
19 & 8 & 668 & 672 & 1340 
\end{tabular}
\caption{
Three-dimensional Ising model: the symmetry of an isotropic square lattice
with periodic boundary conditions combined with the
$Z_2$ color symmetry. The notations are the same
as in Table \protect\ref{t:Ising2d}.
}
\label{t:Ising3d}
\end{table}

\begin{table}
\begin{tabular}{ccccrrrr}
\multicolumn{8}{c}{Three-State Potts Model on a Periodic Square Lattice}\\ 
\multicolumn{8}{c}{N=11 \quad dim=177147} \\ \hline
 R & k & $\lambda$ & $l_R$ & $a_{R,C=0}$ & $a_{R,C=1}$ & $a_{R,C=2}$ & $a_{R}$ 
\\ \hline
0 & 0         &  1 & 1 & 1464 & 1342 & 2806 & 8418\\
1 & $\pi$     &$-1$& 1 & 1221 & 1342 & 2563 & 7689\\
2 & $\pi/11$   & *  & 2 & 2684 & 2684 & 5368 & 16104\\
3 & $2\pi/11$ & *  & 2 & 2684 & 2684 & 5368 & 16104\\
4 & $3\pi/11$ & *  & 2 & 2684 & 2684 & 5368 & 16104\\
5 & $4\pi/11$ & *  & 2 & 2684 & 2684 & 5368 & 16104\\
6 & $5\pi/11$ & *  & 2 & 2684 & 2684 & 5368 & 16104
\end{tabular}
\caption{Three-state Potts model: the symmetries of a periodic chain 
combined with the $S_3$ color symmetry. Same notations as in Table
\protect\ref{t:Ising2d} but $C$ has three possible values
 $C=0$, 1 or 2. There is an extra twofold degeneracy for $C=2$.}
\label{t:Potts}
\end{table}


\twocolumn \input{psfig.tex}

\end{document}